**Estimation of spatio-temporal temperature evolution during laser spot melting using in-situ dynamic x-ray radiography**

Rakesh R. Kamath[1,*], Hahn Choo[1], Kamel Fezzaa[2], and Sudarsanam Suresh Babu[1,3,*]

1. *Materials Science and Engineering, University of Tennessee, Knoxville, TN 37996, USA*

2. *X-ray Science Division, Argonne National Laboratory, Lemont, IL 60439, USA*

3. *Mechanical, Aerospace and Biomedical Engineering, University of Tennessee, Knoxville, TN 37996, USA*

**\* Corresponding authors:**

Rakesh R. Kamath, Ph.D.

The University of Tennessee

Department of Materials Science and Engineering

306 Ferris Hall

Knoxville, TN 37996-2100

Email: rkamath@vols.utk.edu

Sudarsanam Suresh Babu, Ph.D.

The University of Tennessee

Department of Mechanical, Aerospace and Biomedical Engineering

407 Dougherty Engineering Building,

Knoxville, TN 37996-2100

Email: sbabu@utk.edu

**Abstract**

Understanding the spatio-temporal evolution of thermal gradient (G) at and velocity (R) of the solid-liquid and liquid-vapor interfaces is critical for the control of site-specific microstructures in additive manufacturing. In-situ dynamic x-ray radiography (DXR) has been used in recent years to probe the evolution of R with high spatial and temporal resolutions. However, the current methods used to measure the temperature (and therefore, G) are inadequate for sub-melt-pool surface measurement (e.g., thermography) or have lower resolution or limited field-of-view (e.g., x-ray diffraction). In this study, we demonstrate a novel approach to estimate the sub-surface temperature distribution and its time evolution with significantly higher resolutions using DXR data. This methodology uses the Beer-Lambert's law as a physical basis and is demonstrated using an in-situ laser spot-melting experiment on Ti-6Al-4V alloy.

**Keywords (5):** in-situ x-ray radiography, spatio-temporal temperature evolution, Beer-Lambert's law, laser melting, additive manufacturing

Control of microstructure and elimination of pores are essential to obtain desired performance of metallic components produced via additive manufacturing (AM). While the thermal gradient (G) at and velocity (R) of the liquid-solid interface dictate the microstructure [1, 2], the instabilities at the keyhole interface between liquid and vapor as well as the fluid flow affect the formation/retention of the pores [3, 4]. Recently, in-situ dynamic x-ray radiography (DXR) [5, 6] has been used to unravel fundamental mechanisms governing the vapor cavity and melt-pool dynamics. However, the temperature distribution around the vapor cavity and melt-pool has been typically estimated using coupled thermal-fluid flow simulations [3, 7, 8].

Numerical simulations of laser-metal interactions and melt-pool dynamics employ varying degrees of detail in modelling the physical phenomena to estimate the spatio-temporal temperature evolution during both melting and solidification, including heat conduction-only [7], both heat and mass flow enabled [8], and laser absorption via ray-tracing and free-surface tracking [3]. This spatio-temporal evolution of temperature is subsequently used to calculate R (velocity of the liquidus isotherm) and G (temperature gradient at the liquidus isotherm) [9]. A major limitation of these approaches is that they require calibration of the heat source characteristics (i.e., absorptivity), which is often performed using the maximum melt-pool dimensions measured by post-mortem metallography. A better way to calibrate and validate these simulations would be to use transient melt-pool dimensional and temperature evolution measured by in-situ experimental techniques.

To this end, in-situ methods such as synchrotron x-ray diffraction [10-15] and thermal imaging [13, 16, 17] have been used to measure spatio-temporal evolution of temperature during melting and solidification events. The key characteristics of these experimental techniques include: (i) spatial resolution ($x_{res}$), (ii) temporal resolution ($t_{res}$), and (iii) field-of-view (FOV). Typical values of these attributes in previous studies using the x-ray diffraction and thermal imaging are compared to our current approach using the x-ray radiography in Table 1. In the case of x-ray diffraction studies [10-15], a sub-millimeter sized x-ray beam is pointed at a fixed location for the duration of melting/solidification events and transmission-mode diffraction patterns are recorded in sub-millisecond intervals. The changes in lattice parameters extracted from the Debye-Scherrer rings, along with the thermal expansion coefficient as a function of temperature, are used to estimate the time- and volume-averaged temperature from the scattering volume during each time frame. While recent efforts have improved the $t_{res}$ (using higher frequency detectors) [14] and $x_{res}$ (using finer beam size) [15] compared to the previous studies [10-13], the measurement FOV is still limited to a single fixed location in space experiencing changing thermal conditions. Therefore, obtaining a measurement of the evolution of full-field (i.e., entire melt pool and vicinity) temperature distributions as a function of time is a challenging task for the in-situ x-ray diffraction method.

In the case of thermal imaging, the camera captures the infrared radiation originating from the measured FOV, while filtering out the laser reflections. Typically, one or two of the above three attributes need to be compromised to probe the temperature field during in-situ melting/solidification studies. For example, a measurement with a high $t_{res}$ requires an FOV reduction for a similar $x_{res}$ [16] or a high $x_{res}$ demands a decrease in $t_{res}$ and FOV [17]. Even when these characteristics are optimized to a certain extent, a major shortcoming of the thermal imaging techniques is that the FOV is limited to the top surface area of the melt-pool.

In the current study, we propose a method which uses in-situ x-ray radiography imaging to assess the temperature field taking advantage of a combination of high $x_{res}$ of 4 μm, high $t_{res}$ of a few μs, and large sub-surface FOV of the order of mm$^2$ (Table 1). While radiography detectors are generally bound by trade-offs similar to thermal imaging cameras, the relatively high spatial and temporal resolutions employed in the present study still allowed for an FOV sufficient to obtain a simultaneous full-field capture of the entire sub-surface melt-pool dynamics (and consequently, temperature evolution through our methodology). In this paper, we outline a novel approach to estimate the transient, sub-surface temperature distributions from dynamic x-ray radiography images, using the Beer-Lambert's law, measured during laser melting of Ti-6Al-4V alloy. The methodology is explained step-by-step using a case study of a conduction-mode spot melt on a Ti-6Al-4V alloy plate.

The fundamental basis for radiography [5, 6, 18] and tomography [19, 20] tools used in materials science is image contrast based on differences in x-ray mass-absorption. The x-ray mass-absorption is dictated by the atomic number of the elements present in the sampling volume. Irrespective of the type of incident radiation, the measured (transmitted) intensity (I) and incident intensity ($I_0$) are related through the Beer-Lambert's law [21],

$$I = I_0 \exp\left(-\frac{\mu}{\rho} * t * \rho\right) \ldots \ldots (1)$$

where, ($\mu/\rho$) is the mass attenuation coefficient, $\rho$ is the density and $t$ is the thickness of the absorbing medium. In the present case, the data measured in the form of x-ray radiographs are pixels whose grayscale values are the transmitted intensity (I) in Eqn. 1. These values can be used to obtain the density (ρ) corresponding to each pixel using Eqn. 1. In turn, since the density (ρ) is a function of temperature, we can estimate the temperature distributions from the radiography images. Previously, this approach has been applied in computed tomography as reviewed by Fani et al. [22]. While very useful, this method only provides the through-thickness averaged temperature due to the nature of the radiography measurement (i.e., no information in the direction of the x-ray beam/thickness direction). Our current approach addresses this problem and resolves a true 2-D radial distribution of the temperature from the through-thickness averaged temperature map, by using a recipe outlined by Gilbert and Deinert [23] that was originally developed to extract axisymmetric flows in soils measured using neutron radiography data.

A conduction-mode spot melt was produced on a Ti-6Al-4V (Ti-64) plate in an argon atmosphere using a laser power of 82 W and dwell time of 8 ms. Simultaneously, DXR (with spatial and time resolutions of 4 μm and 14.28 μs, respectively) was used to acquire a time sequence of 741 images over 10 ms during the melting and solidification cycles. More details on the experimental setup are provided in [18]. The workflow illustrating various steps from an unprocessed radiography image to a temperature map is presented in Fig. 1. Each step in the workflow is implemented using the MATLAB code written in-house and is available upon request with the authors.

*Step 1 - Filtering and background correction of unprocessed DXR images:* An example of as-measured 2-D radiography image shows argon atmosphere (top, yellow) and Ti-64 plate (bottom, blue), Fig. 1a. First, the top part of the FOV consisting of the argon atmosphere is removed for ease of subsequent analysis. The bottom part of the images consisting only of the Ti-64 plate is then processed using (i) median filtering and (ii) background correction. The median filtering process replaces each pixel with the median of the 8 nearest neighbor pixels around it, which is commonly used in image processing for speckle removal. The background correction step consists of two sub-steps, (i) instrument correction and (ii) time-series correction. The first sub-step involves the construction of a correction matrix by dividing the median of the topmost row of pixels (which is assumed to correspond to room-temperature Ti-64) by each pixel intensity in the first image. This is followed by multiplication of each image in the time sequence by the correction matrix. This sub-step is used to separate the effect of systematic factors on measured intensity such as contrast gradient within a given image due to substrate thickness variations, detector sensitivity, and variation in incident intensity (due to detector assembly) in the FOV. The second sub-step consists of dividing the instrument corrected pixel arrays in each timestep by the median of a given row of pixels and multiplying the median of same row from the first image. This sub-step ensures correction of the steady increase in detected intensity as a function of measurement time (shown in Supplementary Fig. S1). The corrected intensity map is shown in Fig. 1b. Note that, from this step onwards, only right half of the melt-pool region (with full depth, indicated by the white dotted box in Fig. 1a) is analyzed for ease of processing. Therefore, the zero (origin) of the width-depth maps in Figs. 1b-e is approximately the center of the melt-pool.

*Step 2 - Resolving through-thickness intensity:* The background-corrected intensity map is then fit to a 3-D surface using a MATLAB functionality to obtain a "fit intensity" map. Then, the radial distribution of the intensity is extracted from the images using the recipe described in Gilbert and Deinert [23]. The input is the fit intensity map, which is essentially a weighted, through-thickness average of different degrees of absorption due to the relative changes in the path-length of x-ray through solid and liquid at varying temperatures. Using the Gilbert-Deinert approach, the output obtained is a three-dimensional, radially-resolved (i.e., including along the through-thickness direction) intensity distribution. Fig. 1c shows the 2-D intensity map resolved at the mid-thickness of the plate.

*Step 3 - Conversion of radially-resolved intensity maps to density maps:* The time sequence of radially-resolved intensity maps is then converted to density maps using the Beer-Lambert's law (Eqn.1). In this step, the first task is to calibrate the conversion from intensity to density. This is achieved by using the first image in the radiography measurement sequence, which was measured at room temperature. Based on the absorption coefficients at the given x-ray energy used (approximately 25 keV), the Ti-64 region absorbs significantly more x-rays than the argon (Ar) region [21]. For the calibration process, the $I_o$ (= $I_{Ar}$) and $I_{Ti-64}$ are taken to be equal to the median intensity obtained from the top row of pixels in the Ar region and median intensity from the top row of pixels in the Ti-64 plate, respectively, in the first image of the sequence. Further, using room-temperature density values of the Ti-64 (4430 kg/m$^3$) and the Ar (1.652 kg/m$^3$) and the corresponding intensity values, the ($\mu t/\rho$) term in Eqn.1 is determined. The ($\mu t/\rho$) factor can be used to calculate the density corresponding to each intensity point in the image sequence. The result obtained from this step (i.e., a density map) is shown in Fig. 1d. Note that this map is the 2-D density distribution in the mid-thickness of the plate.

*Step 4 - Conversion of density maps to temperature maps*: Finally, the density maps are converted to temperature maps, Fig. 1e. The literature values [24-26] for the variation of densities of both solid and liquid phases of Ti-64 are fit to polynomial functions using the "polyfit" function in MATLAB. Note that the temperature ranges, polynomial fit coefficients, and corresponding references are summarized in Table 2. More details on the density vs. temperature data from the literature and the fit polynomials are presented in Supplementary Figure S2. This step is then repeated for entire time sequence to obtain the temperature maps as a function of time with intervals corresponding to the temporal resolution of the DXR measurement (14.28 μs).

Fig. 2 shows a set of temperature maps selected at key stages during the laser spot melting and solidification. The melting stage includes maps from Fig. 2a (350 μs after laser is on) to Fig. 2e (8000 μs, when the laser is turned off). The solidification stage is depicted from Fig. 2e (8000 μs) to Fig. 2h (8780 μs). As the laser-on dwell time increases (from Figs. 2a to 2e) the center of the melt-pool gets hotter (increasingly red in color) and the liquidus isotherm (1930 K corresponding to the bluish-green tint) moves radially outward indicating an increase in the melt-pool size. After the laser is off, the melt-pool solidifies as can be seen from the shrinking region enveloped by the liquidus isotherm. Additionally, the melt-pool area cools very rapidly, while the heat in the former heat-affected zone (HAZ) moves slowly outward (yellow-green region), suggesting a slower cooling rate in this area.

Taking advantage of the high spatial resolution of the temperature map (4 μm), a series of thermal gradient (G) maps can be obtained as a function of time, which would be useful for the studies of microstructure development during the solidification. Figs. 3a and b show the 2-D temperature map and corresponding thermal gradient (G) map from Frame #50 (i.e., 700 μs into melting), respectively. Figs. 3c and d show the corresponding 1-D profiles for temperature and thermal gradient as a function of depth at various locations along the width axis. Fig. 3b shows that the thermal gradient in the probed region is of the order of $10^6 – 10^8$ K/m. These values are consistent with those observed in spot melt simulations on materials with widely different thermophysical properties over the past few decades by Katayama et al. on aluminum [27], Debroy et al. [9] on steel, and Raghavan et al. on nickel [7]. The value of G is the highest in the melt-pool, especially at the top, which is also observed in modelling studies as the main driving force for Marangoni flow [7-9]. While the thermal gradient at the top-center of the melt pool seems to be zero, it should be higher than the periphery. This is caused by a limit on the density data being unavailable for T > 3560 K for Ti-64. The thermal gradient in the mushy zone is about an order of magnitude lower than the melt-pool, whereas the G in the heat affected zone (HAZ) is similar to that in the melt-pool. However, the hotspots seen in the envelope of the HAZ and edges of the probed region is most likely an artifact of the image processing steps used and the finite boundaries of the sample plate.

The liquidus isotherm (horizontal dotted line) intersecting the temperature profiles shown in Fig. 3c indicates that the melt-pool depth is approximately 100 μm which is somewhat larger than the experimentally measured value (from the DXR images) of 35 μm [18]. One of the main causes for this discrepancy, which will be discussed later, may be related to the scaling operation (i.e., Step 3). Additionally, the lines in Fig. 3c show the presence of a wider mushy zone (region with lower slope below the liquidus isotherm) than reported in previous modelling studies during the melting stage [7-9]. These observations from Fig. 3c indicate that the liquid-solid interphase boundary distinguishable in the x-ray radiography images might correspond to the liquidus isotherm (rather than the solidus isotherm). Fig. 3d shows that the thermal gradient, for example at w = 80 μm, is high at the center of the melt-pool and gradually decreases with increasing depth. As mentioned earlier, the G values near the top-center of the melt pool (w = 0 ~ 60 μm) are zero due to the limited density data as a function of temperature, but in principle should show a similar trend as the profile at w = 80 μm. Further, a simple extrapolation of the G line profiles (corresponding to w = 0 ~ 60 μm) towards the melt-pool center indicates that the G at the top-center of the melt-pool should be higher than its periphery.

The high temporal resolution of the DXR measurement allowed the extraction of temperature maps with a time interval of 14.28 μs between two frames in the present study. These temperature maps for successive timesteps were used to obtain the melt pool dimensions (Fig. 4a) and G at the liquid – mushy zone interface along the depth direction (Fig. 4b) as a function of solidification time. Fig. 4a shows that the melt pool depth decreases from about 120 μm to 0 μm in about 600 μs. These values agree well with the direct analysis of the DXR images of the melt-pool depth, which decreases from about 120 μm to 0 μm in about 700 μs. The fact that our temperature analysis scheme was able to capture the melt-pool solidification kinetics reasonably well provides some validation to our approach. Fig. 4b shows that G is relatively constant at around 2 ~ 4 x $10^7$ K/m for most of the solidification time which is consistent with previous modelling studies [7, 9, 27].

While the temporal resolution remains unaffected by the operations performed in the current workflow, the spatial resolution of the temperature maps is slightly affected by the smoothing/filtering operation. The various contributors to spatial resolution of the measurement are discussed in supplementary section S3. Further, the uncertainty in temperature values is dependent mostly on the uncertainty in experimental density vs. temperature data and slightly on the polynomial fits used for the conversion of density maps to temperature maps. Another source that may result in systematic uncertainty is the scaling operation in Step 3, i.e., the $I_{Ti-64}$ and $I_{Ar}$ values used. This can be overcome by the measurement of a standard sample with varying thicknesses for a given DXR setup and preforming the intensity-density conversion from such calibration measurements.

In summary, the highlight of this work is that the Beer-Lambert's law, along with the Gilbert-Deinert approach, can be used to design a methodology to assess the transient, sub-surface temperature distributions using x-ray radiography images. Further, this methodology was demonstrated using a case study of a conduction-mode spot melting on Ti-6Al-4V alloy. The reasonable estimates of transient G and R obtained as a product of this workflow and DXR measurements can be correlated with the local solidification microstructure to validate/improve solidification theories. It can also be applied to aid the design of AM processes, calibration of in-situ monitoring equipment in AM machines, and to help increase the accuracy of high-fidelity melt-pool simulations. Also, in addition to understanding the transient evolution of thermal gradient (G), the temperature maps can be potentially used to automate the extraction of melt-pool and vapor cavity dimensions (by tracking the liquidus and vaporization temperature isotherms, respectively). The variation of cooling rate as a function of solidification time can also be extracted and be used to develop or validate existing empirical relations to calculate microstructure feature sizes during solidification. Further, the temperature maps can also be converted into liquid fraction maps, thus help us better understand the behavior of mushy zone during both melting and solidification.

**Acknowledgements**

This research was sponsored by the Department of the Navy, Office of Naval Research under ONR award number N00014-18-1-2794. Any opinions, findings, and conclusions or recommendations expressed in this material are those of the author(s) and do not necessarily reflect the views of the Office of Naval Research. This research used resources of the Advanced Photon Source, a U.S. Department of Energy (DOE) Office of Science User Facility operated for the DOE Office of Science by Argonne National Laboratory under Contract No. DE-AC02-06CH11357. This work was initiated as a part of the Ph.D. dissertation study (made available online in Aug 2022) [18] of R. R. Kamath under the supervision of Prof. H. Choo at the University of Tennessee Knoxville.

**Table 1**. Various in-situ experimental techniques used to extract the melt-pool temperature with their corresponding parameters such as field of view (FOV), spatial ($x_{res}$) and temporal resolutions ($t_{res}$).

| Technique | Spatial resolution (μm²) | Temporal resolution (μs) | Field of view | Ref. |
|---|---|---|---|---|
| X-ray diffraction | 300 x 50 | 1000 | Sub-surface | [10] |
| | 100 x 40 | 4000 | Sub-surface | [11] |
| | 100 x 100 | $6.67 \times 10^6$ | Sub-surface | [12] |
| | 30 x 9 | 1000 | Sub-surface | [13] |
| | 140 x 80 | 50 | Sub-surface | [14] |
| | 110 x 30 | 50 | Sub-surface | [15] |
| Thermal Imaging | Not specified | 2000 | Only surface | [13] |
| | 30 x 30 | 100 | Only surface | [16] |
| | 30 x 30 | 10 | Only surface | [16] |
| | 30 x 30 | 50 | Only surface | [17] |
| | 4 x 4 | 1000 | Only surface | [17] |
| X-ray radiography | 4 x 4 | 14.28 | Sub-surface | Current study |

**Table 2**. The coefficients of the fit used for experimentally measured values of densities as a function of temperature for different phases of Ti-6Al-4V and corresponding sources. The functional form of the fit used is $\rho = aT^2 + bT + c$.

| Phase | T range (K) | Fit coefficients | Ref. |
|---|---|---|---|
| Solid | 270 - 1870 | a = -6.1558 x $10^{-5}$, b = 0.0245, c = 4428.2 | [24] |
| Mushy | 1870 - 1920 | a = 0, b = -2.5415, c = 9011.9 | [24],[26] * |
| Liquid | 1920 - 3560 | a = 0, b = -0.53, c = 5142.4 | [24],[25] |
| Vapor | 3560 - | - | [26] |

* Only mushy zone T range

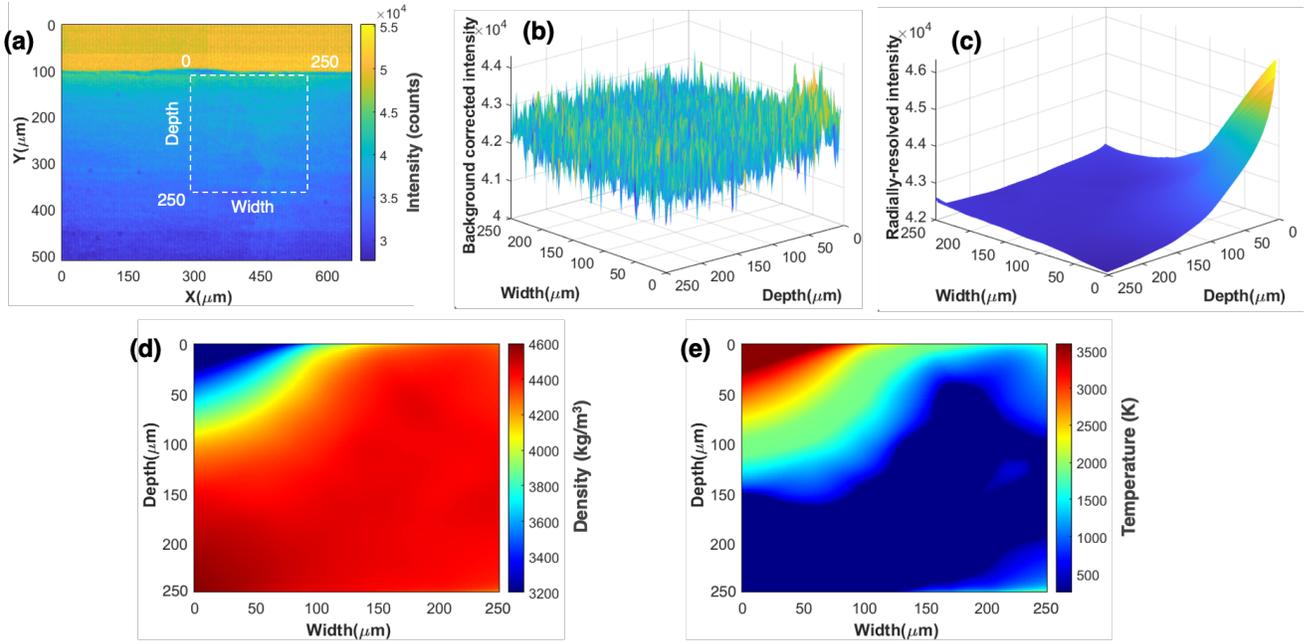

**Fig. 1.** Analysis scheme used to convert the as-measured radiography image to a temperature map. Representative outputs for each step of the analysis pipeline are shown using a timestep of 700 μs (frame #50 from laser on) as an example. (a) The raw radiography image with the argon atmosphere in yellow and Ti64 substrate in blue. A 250 μm x 250 μm area containing half of the melt pool (white dotted box) is used for further analysis. The approximate center of the melt pool is indicated using a 0 on the white box along with the new width-depth reference frame. (b) After step 1: Filtered and background-corrected pixel array for the white boxed area. (c) After step 2: Radially-resolved intensity pixel array. (d) After step 3: Density map obtained from the conversion of the radially-resolved intensity map using the Beer-Lambert's law. (e) After step 4: temperature map obtained from the density map.

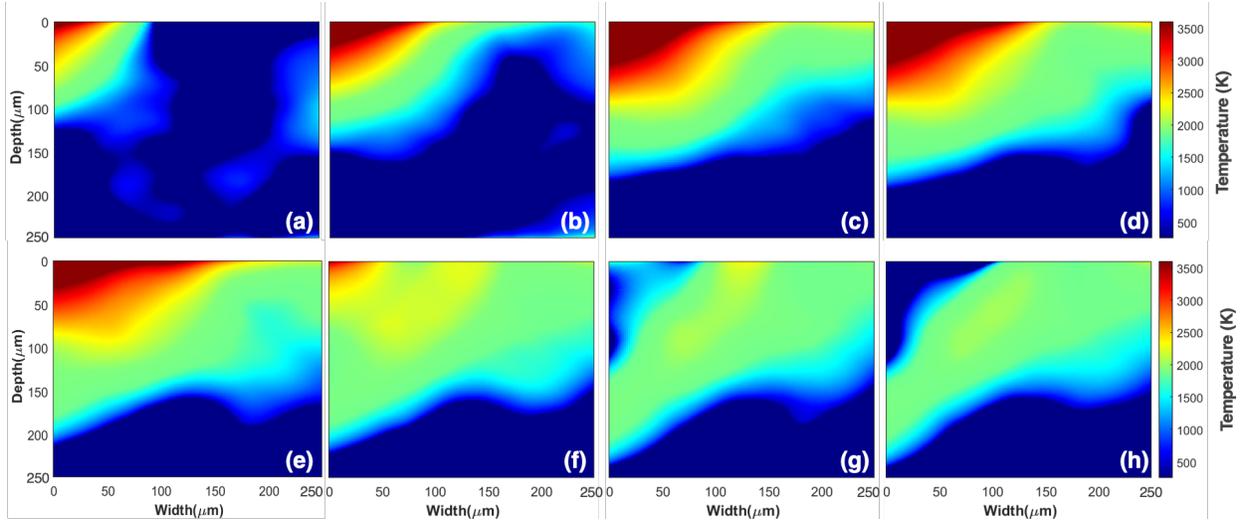

**Fig. 2**. A series of 2D temperature contour maps around the melt-pool region at salient timesteps during the laser spot melting and solidification. (a) 350 μs after laser is on (frame # 25), (b) 700 μs (frame # 50), (c) 2500 μs (frame # 175), (d) 4000 μs (frame # 280), (e) 8000 μs (frame # 560, laser is off and solidification starts), (f) 8350 μs (frame # 585), (g) 8640 μs (frame # 605), and (h) 8780 μs (frame # 615, solidification is complete according to the DXR image). The color scale bar shows temperature in K. The top left corner of each map corresponds to the center of the laser and the melt-pool.

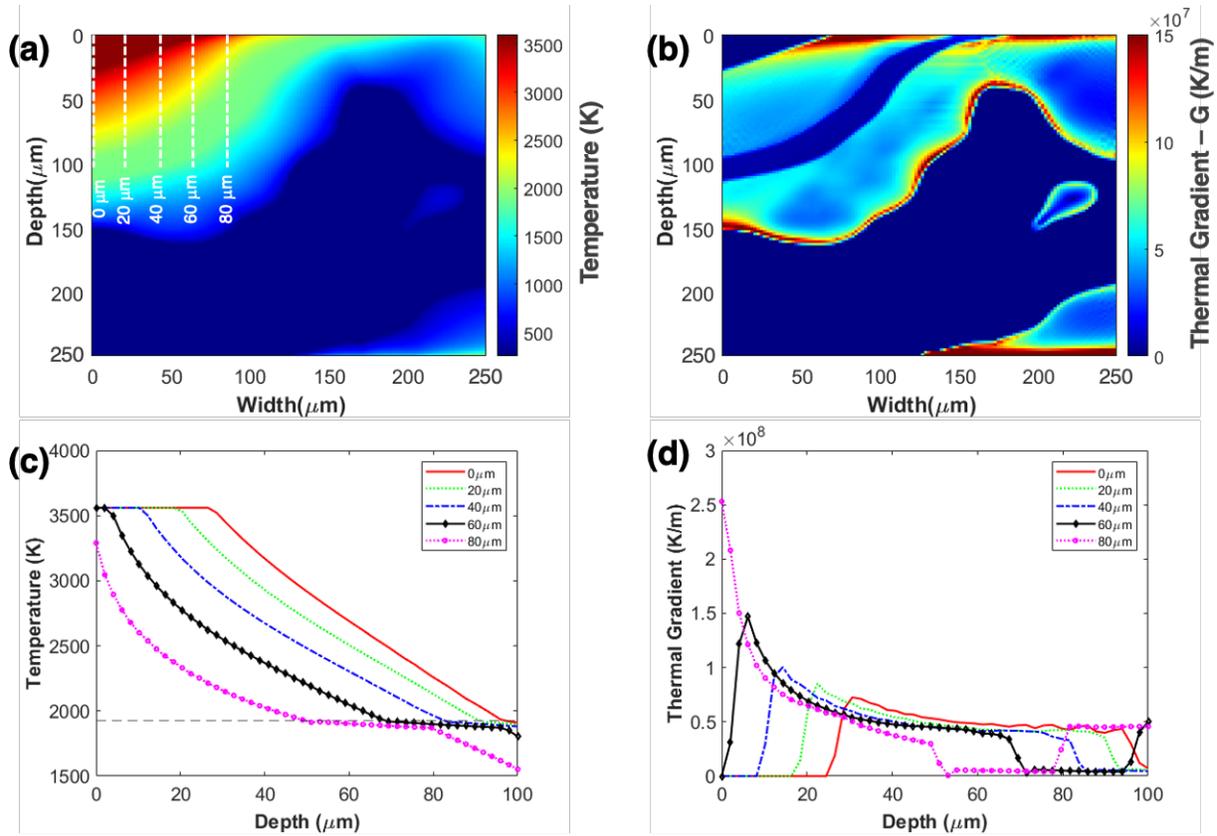

**Fig. 3**. The thermal gradient (G) map obtained from the temperature map. (a) Temperature map at 700 μs from laser on, (b) corresponding thermal gradient (G) map, (c) 1-D temperature profile along the melt-pool depth up to 80 μm at various locations along the melt-pool width from 0, 20, 40, 60, and 80 μm from the melt-pool center (as marked in (a) using white dotted lines). Horizontal dashed line indicates the liquidus temperature (1930 K). (d) corresponding variation of thermal gradient along the depth for the same locations along the melt-pool width. Note that the thermal gradients shown here are normal to the 3-D temperature surface / 2-D temperature contours and therefore, to the liquid – mushy zone interface (i.e., the liquidus temperature contour). Also note that G values in (b) and (d) are calculated using a 2 μm pixel size and therefore, depict two times the value (with the same trend) for a 4 μm pixel size.

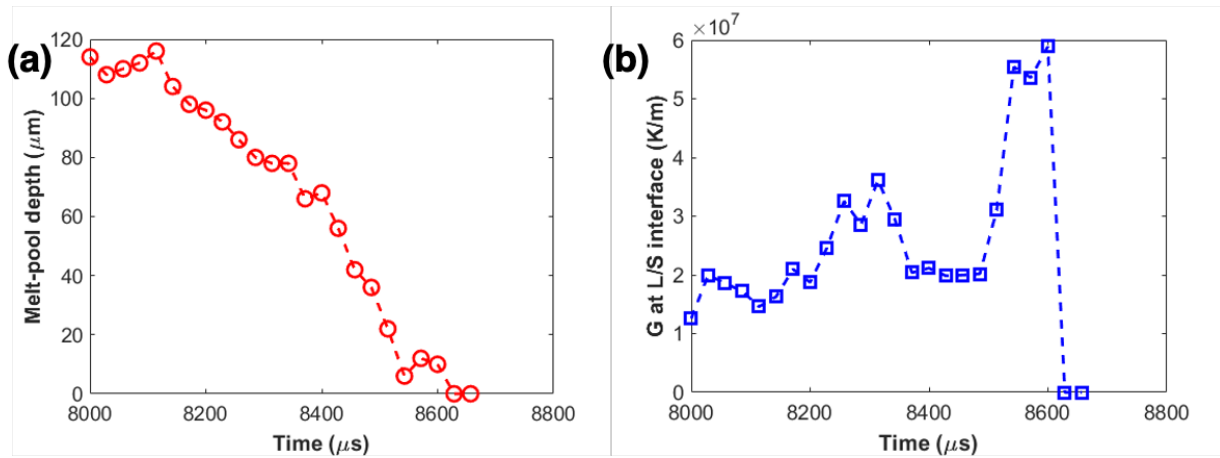

**Fig. 4**. (a) Changes in melt pool depth during solidification as a function of time (laser off at 8000 µs). (b) Changes in thermal gradient (G) at the liquid - mushy zone (L/S) interface along the depth direction (at w = 0 µm) during solidification as a function of time after laser off. Note that G values are calculated using a 2 µm pixel size and therefore, depict two times the value (with the same trend) for a 4 µm pixel size.

# Supplementary information

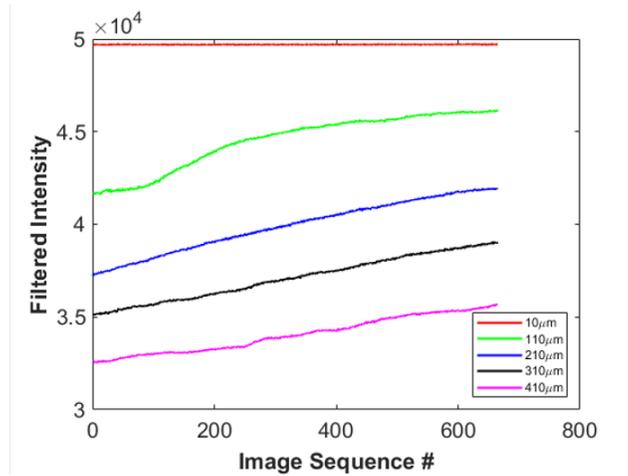

**Supplementary Fig. S1** The median intensity of rows corresponding to Y = 10 (argon), 110, 210, 310, and 410 μm (bottom part of Ti-64 substrate), in Fig. 1a, as a function of image sequence number. Note that image sequence #1 corresponds to 0 μs (i.e., laser on) and image sequence #560 corresponds to 8000 μs (i.e., laser off). The figure shows that median intensity of all the rows increases with time. However, a basic thermal diffusion calculation tells us that for the given laser-on time frame (8 ms), the lower parts of the Ti-64 substrate (Y = 410 μm) should not experience any heating. This justifies the need for the time-series correction sub-step implemented as a part of the background correction in Step 1 of the workflow.

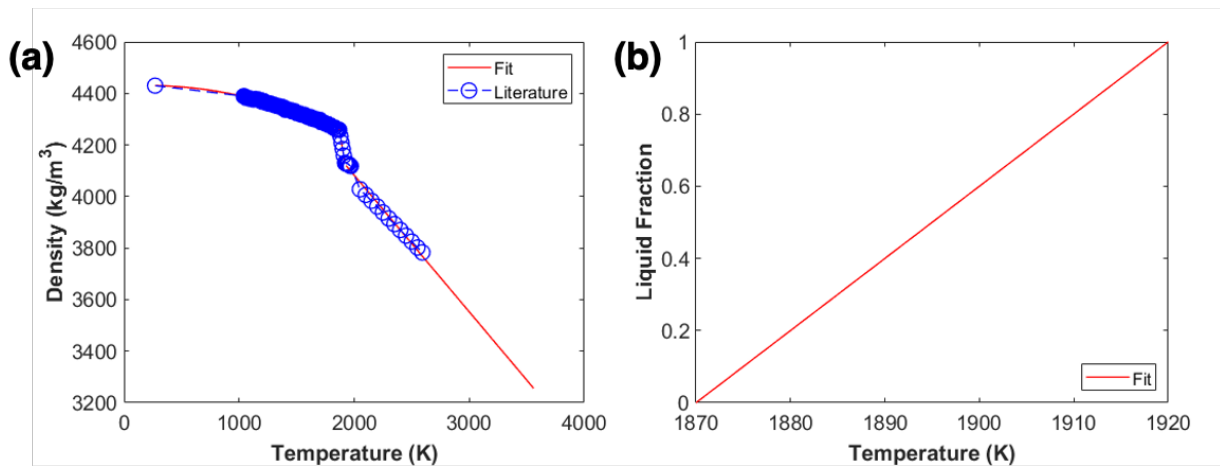

**Supplementary Fig. S2** (a) Density vs temperature data obtained from the literature [24-26]. The assumption made here during the calculation of the density of the solid and liquid mixture in the mushy zone is that the liquid fraction varies linearly with the temperature. This is shown as a plot of liquid fraction vs temperature in the mushy zone in (b).

**Supplementary section S3 - Contributors to spatial resolution of the measurement**

The raw imaging data obtained contained square pixel with dimensions of 2 μm each. However, the effective spatial resolution of the square pixels was determined to be about 4 μm and the contributors to the same have been discussed in detail below.

*Smoothing step*

The median filtering operation, performed as a part of Step 1, assigns the running average of 8 neighboring pixels to a given pixel. This effectively changes the spatial resolution from 2 μm (1 pixel) to about 4 μm (2 pixels).

*Point spread function of the detector*

The detector used is a Photron SA-Z high-speed camera with 1024 x 1024 pixel resolution with each pixel being 20 μm in size. A Mitutoyo 10X Microscope objective (NA 0.28) is used to image the 100 μm thick LuAG:Ce scintillator crystal in the camera providing an effective pixel size of 2 μm for the measurement. The point spread function (PSF) of this detector was measured by scanning a knife edge (GaAs cleaved crystal) positioned against the scintillator. The scan steps (0.2 μm) were much smaller than the effective pixel size (2 μm). A gaussian fit to the measured peak yield a 4 μm (FWHM) for the PSF.

*Finite size of X-ray source*

Another contribution to the overall spatial resolution comes from the finite size of the x-ray source (282 μm FWHM). Given the source-sample distance (38 m) and the sample-detector distance (0.35 m), the projected horizontal source turns out to be 2.6 μm FWHM. Adding in quadrature the two contributions yields a 4.8 μm FWHM total detection resolution.

*Sample expansion due to heating by X-ray beam*

The power of the beam incident on the 1 x 1 mm$^2$ area (approximately equal to the FOV used in the study) at the sample location (38 m from the source) is about 17 W. For the sample size used (50 mm x 3 mm x 0.5 mm) this has seen to result only in the expansion of the about 2-4 μm during the 8 ms duration (without the laser source switched on). Also, it is reasonable to assume that the expansion in the perpendicular direction (along x-ray beam) should also be in a similar (2-4 μm) and doesn't change the x-ray path length significantly enough contribute to the accuracy of the temperature measurement.